# A Practical Flake Segmentation and Indexing Pipeline for Automated 2D Material Stacking

Yutao Li[1#], Logan Sherlock[1], Ryan Benderson[1], Daniel Ostrom[2], Huandong Chen[1], Kazuhiro Fujita[1#], Abhay Pasupathy[1,2#]

1. Department of Condensed Matter Physics and Materials Science, Brookhaven National Laboratory, Upton NY, 11973
2. Department of Physics, Columbia University, New York, NY, 10027

#. Corresponding author

**Abstract:**

A cost-effective and robust image-processing pipeline is presented for the detection and characterization of exfoliated two-dimensional (2D) material flakes in optical microscope images, designed to facilitate automation in van der Waals heterostructure assembly. The system combines shallow machine learning (ML)–based material classification with a precision-first flake detection algorithm driven by edge morphology and color discontinuity. Step edges are resolved when supported by optical contrast, while spurious features such as dust and background texture are reliably rejected. Each identified flake is exported in a structured format that includes centroid coordinates, bounding geometries, average RGB color, and estimated optical thickness, enabling seamless integration into automated pick-up and stacking workflows. The pipeline is hardware-light and operates without the need for deep learning models or nanoscale ground-truth labels, making it practical for scalable front-end wafer processing at a hardware cost of under $30,000. In contrast to prior approaches that focus solely on detection accuracy, the proposed system unifies flake segmentation with indexing, filtering, and blueprint-driven stacking, forming a closed-loop workflow from image acquisition to device planning. Its low annotation requirement and flexible implementation enable rapid deployment across diverse 2D material systems and imaging conditions.

**Introduction:**

Mechanically exfoliated two-dimensional (2D) materials, such as graphene, hBN, and transition metal dichalcogenides (TMDs), remain essential for research-scale device fabrication due to their high crystallinity, clean interfaces, and tunability in custom heterostructures. However, the process of locating and selecting suitable flakes still relies heavily on manual inspection under an optical microscope—a major bottleneck for repeatability, throughput, and automation. The challenge arises from the irregular geometry, varying optical contrast, and fragile morphology of exfoliated flakes, which complicate both classification and boundary segmentation. These issues are further exacerbated by artifacts such as dust, silicon debris, and spatially non-uniform illumination. [1–9]

Existing methods for automated flake detection have ranged from deep convolutional networks to classical thresholding and color filtering. While many demonstrate promising results on small, curated datasets, most depend on extensive ground-truth labeling, struggle with false positives from contamination, or lack robustness across imaging conditions. Some recent works have applied unsupervised methods such as k-Means clustering to characterize internal domains within pre-selected flakes, but have not addressed flake detection itself or integration into an end-to-end workflow. [10–13]

In this work, we present a modular pipeline for reliable flake classification and segmentation directly from optical microscope images acquired under controlled illumination. The system combines shallow machine learning for material type classification with an edge-aware flake detection module based on Canny filtering and k-Means clustering. Unlike prior pipelines that assume flakes are already localized, our method performs full-field flake identification and outputs structured data—flake boundaries, flake-subflake relation hierarchy, average RGB, and estimated thickness—suitable for downstream automation such as robotic pickup and stacking workflows. It requires no deep learning, nanometric ground truth, or hyperspectral input, enabling practical deployment on real wafer-scale devices. [14–18]

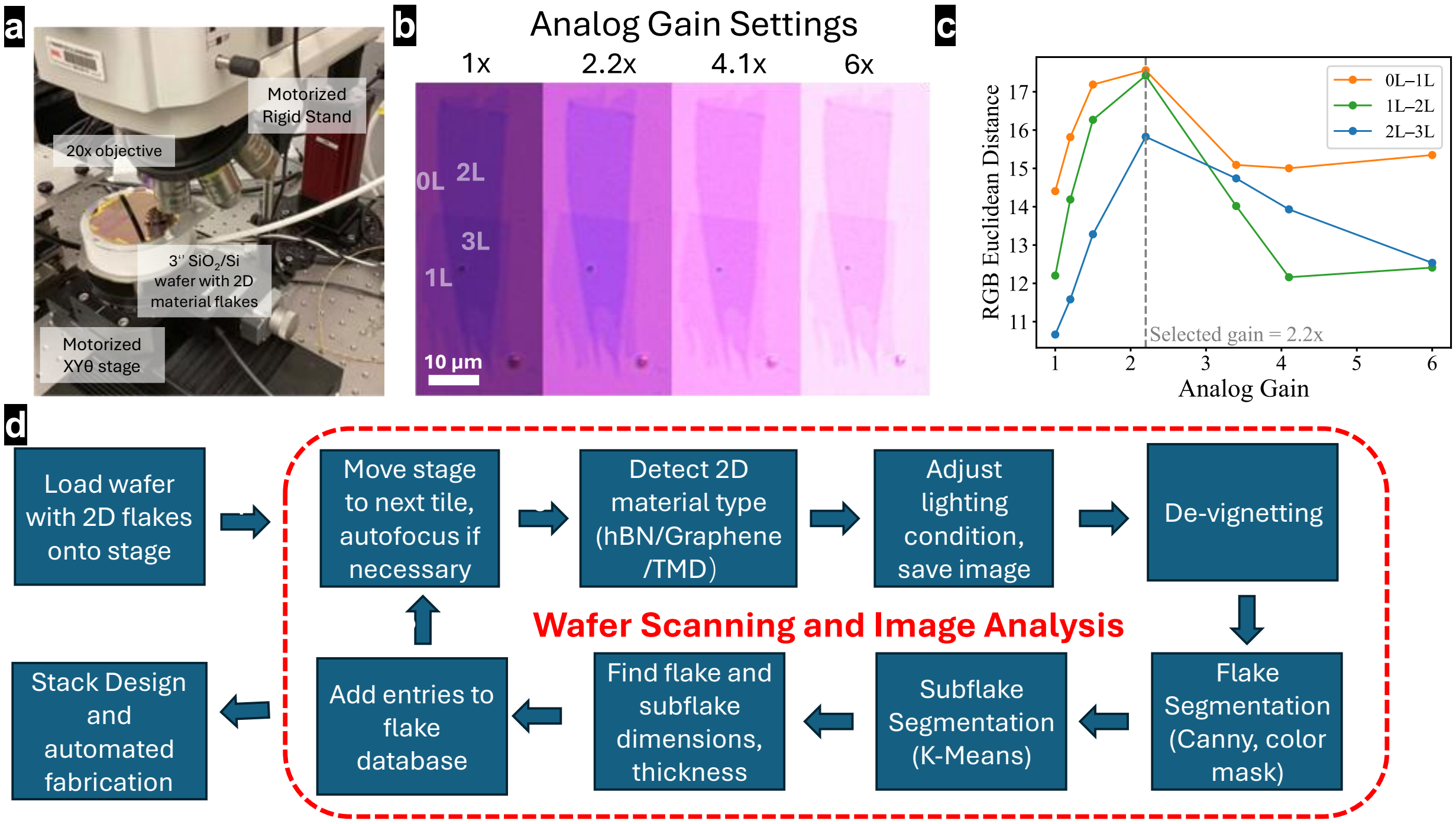

**Figure 1.** *Overview of the 2D material flake cataloging pipeline.*

(a) Schematic of the hardware setup for image acquisition. The system consists of 4 stepper motors that can be controlled by Python code: XYΘ for the stage, and Z for the microscope head. The objective is stationary in XY directions. Image acquisition across the entire 3'' wafer is done by rastering the stage in XY.

(b) Optical microscopy images of the same graphene flake on a 285 nm $SiO_2$/Si wafer, captured at four different Analog Gain settings (exposure time = 100 ms). Regions corresponding to 0–3 atomic layers of graphene are labeled.

(c) Euclidean RGB distance between adjacent layer regions (0L–1L, 1L–2L, 2L–3L) as a function of Analog Gain (exposure time = 100 ms), indicating that 2.2× gain provides optimal color contrast.

(d) Flow diagram of the flake cataloging pipeline.

### Image Acquisition and Classifications

Reliable image-based flake detection begins with high-quality input. When optical images are sharply focused and lighting conditions are tuned to maximize color contrast—both between flakes and the substrate, and between regions within a single flake differing in thickness by only a few atomic layers—subsequent classification and segmentation become significantly more accurate and robust. In this work, we define such intra-flake regions as subflakes, which typically correspond to discrete thickness domains separated by step edges. Rather than compensating for poor imaging with post-processing or complex models, our pipeline emphasizes clean, high-contrast image acquisition at the source. [14–18]

Optical images were acquired on a Nikon LV150N reflected-light microscope equipped with a 20 × objective and a halogen lamp operated at full power (manufacturer-specified CCT ≈ 3400 K) (Fig. 1a). Exfoliated 2D materials are situated on a 3" Si wafer coated with a 285 nm wet thermal oxide layer. The wafer is held in place by vacuum suction on a 3.5" motorized stage, and its location is controlled by XYΘ stepper motors. After the wafer is loaded onto the stage, it is automatically aligned so that its [110] crystal axis is aligned to the X axis of the stage motor. Then the entire wafer is scanned by rastering the XY stage across the wafer in a grid pattern. Since the wafer surface is not perfectly flat, automated refocusing, physically enabled by the Z motor on the microscope head stand, is performed at 7 mm intervals using a focus metric based on the high-frequency components of the image's green channel, computed via fast Fourier Transform (FFT). This enables consistent sharpness throughout large-area scans.

Different 2D materials require different lighting conditions for optimal contrast. For example, a 100 ms exposure with 2.2× analog gain provides sufficient contrast to differentiate monolayer, bilayer, trilayer, 4-layer, and thicker graphene from each other and from the $SiO_2$ background. In contrast, moderately thick hBN flakes (~50 nm), which are commonly used as dielectric layers, are best imaged under 50 ms exposure and 1× gain. As shown in Fig. 1b-c, inappropriate lighting settings significantly reduce flake visibility and step-edge contrast. [14–17]

Because our system enables mixed-material exfoliation on a single wafer—greatly simplifying downstream automated 2D material device fabrication—each newly scanned image is immediately processed by a lightweight machine learning model trained on manually annotated 2D material type from both lighting setups. While the CNN operates as a black box, interpretable visual features correlate with each class: BN flakes typically appear more colorful; graphene flakes often contain bluish or purplish segments; and TMD flakes tend to show uniform yellow tones, associated with >100 nm thickness flakes that make up the majority of the exfoliated flakes. The model classifies the field of view as containing graphene, hBN, TMD, or none. If the region is classified as "none" or the confidence level is below a certain threshold the image is discarded. Otherwise, the system switches to the optimal imaging parameters for the identified flake type and re-acquires the image. (Supplementary Information 2)

The model runs at approximately 20 ms per image and achieves over 98% classification accuracy, with no observed confusion between graphene, hBN, and TMD categories (Fig. S1). Color variation due to wafer-to-wafer differences (e.g., batch-dependent $SiO_2$ optical properties) is addressed via minor adjustments to the camera's white balance settings. After approximately 8 hours of operation, the system scans ~16,000 images (each 640 μm × 440 μm field of view) across a full 3" wafer.

**Flake Segmentation from Optical Images**

Our algorithm mimics the way humans identify and catalog 2D material flakes from optical microscope images. First, candidate features resembling exfoliated flakes (instead of tape residue or dust particles) are identified — this is referred to as flake segmentation. Second, each flake is further analyzed to distinguish internal regions of different thicknesses. Since optical contrast and color on Si/$SiO_2$ substrates correlate strongly with thickness, these subflakes can be segmented based on subtle variations in color — a process we term subflake segmentation. This two-step approach reflects how researchers manually inspect optical images when selecting flakes for device fabrication. [13–15]

We utilize different strategies for hBN and graphene flake segmentations. Most 2D material-based devices utilize hBN flakes with thickness between 2nm (violet) to 100nm (red) as dielectric interlayers. Therefore, under the lighting condition described above, these flakes, whose color span most of the visible spectrum, maintain significant contrast with the SiO2 background. [7]

Once an optical image has been classified as containing hBN by the machine learning model, it undergoes Gaussian blurring to suppress noise, and the red channel is extracted. Canny edge detection is then applied to the quantity $2|r - 49|$, as the optical contrast between thin (~5 nm) hBN flakes and the $SiO_2$ background is most pronounced in the red channel. The resulting edges are dilated and subsequently eroded to close small gaps.

These edges — introduced by flakes, tape residues, and dust particles — partition the image into disjoint regions. Regions smaller than one-third of the image size (to exclude the background) and larger than a minimum area threshold (~20 $\mu m^2$) are retained as flake candidates. Since some, but not all, internal step edges may already have been detected by the Canny edge detection algorithm, we dilate all candidate regions to ensure that segments belonging to the same flake are merged. The area surrounding each merged region is then cropped and saved for subflake analysis.

Importantly, non-flake features such as tape residues and dust typically contain internal structures that generate numerous fine edges. These lead to smaller inter-edge regions that often fall below the size threshold. As a result, most of these artifacts are automatically filtered out at this stage.

The segmentation of few-layer graphene flakes requires a different approach. In 2D materials research, graphene serves both as the primary subject of investigation (typically 1–6 atomic layers) and as a conductive gate (5–20 nm). Thin graphene layers, especially those close to the $SiO_2$ background in color, are particularly difficult to segment using edge-based methods such as the Canny edge detection algorithm. However, for flakes thinner than ~20 nm, the optical appearance consistently falls within a black-to-violet range. This enables a color-based segmentation strategy, which we adopt to identify such graphene flakes. [9,14,15,17]

Given that the $SiO_2$ background has a red channel value around R = 204, while monolayer graphene appears closer to R = 188, we select—on the devignetted image (see Supplementary Information S1)—pixels satisfying **45 < R < 198** and **G < 140**. This color threshold is designed to exclude extremely bright features such as tape residue, and extremely dark features such as dust particles. However, the edges of such non-flake features often contain intermediate pixel values that fall within the selection range, creating "webs" of selected pixels that can spuriously connect distinct flakes. To suppress these artifacts, we apply morphological erosion to break such webs.

The remaining pixels are grouped into connected regions and subjected to four geometric filters:

1. **Minimum area (> 50 μm$^2$)** — to exclude noise and extremely small features.
2. **Complexity (< 25)** — defined as the perimeter divided by the square root of the area; this penalizes highly irregular shapes (e.g., dust) while preserving compact flake regions.
3. **Solidity (> 0.4)** — the ratio of region area to its convex hull area; this eliminates highly concave or fragmented shapes.
4. **Extent (> 0.2)** — the ratio of region area to its bounding box area; this further excludes narrow or elongated noise artifacts.

These criteria are intended to isolate genuine graphene flakes while filtering out spurious non-flake features. The area surrounding each retained region is then cropped and saved for subsequent subflake analysis. At this stage, a small number of non-flake features may still pass the filters; these will be further rejected during the subflake segmentation process, which examines internal optical structures more closely.

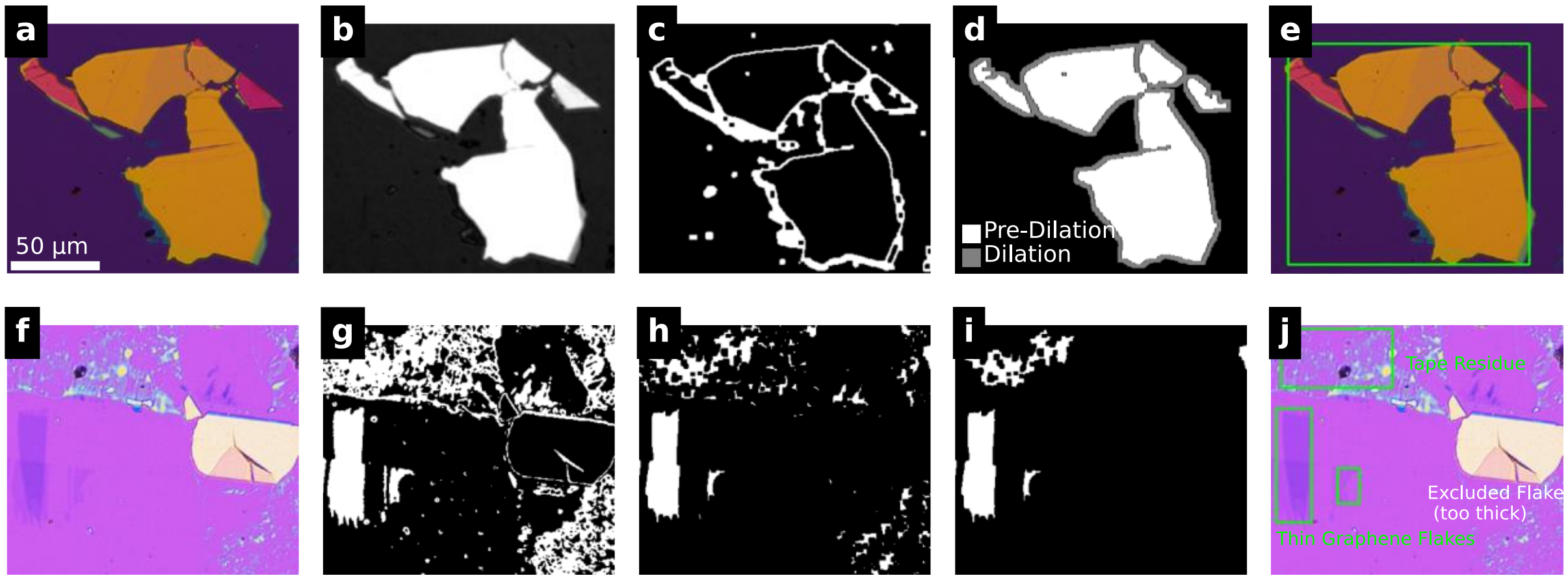

**Figure 2: Flake segmentation.**

(a) Optical microscope image of a $SiO_2$/Si wafer with mechanically exfoliated hBN.

(b) Extracted red channel from a Gaussian-blurred version of (a), plotted as $2 \times |R - 49|$ to enhance flake contrast.

(c) Canny edge detection result from (b), using cv2.Canny(r_diff, 30, 72).

(d) Disjoint regions defined by the edges in (c), shown before and after dilation (background excluded).

(e) Bounding box from (d) applied back to the original image, defining the crop used for subflake segmentation.

(f) De-vignetted optical image of graphene flakes, tape residue, and dust particles on a $SiO_2$/Si wafer.

(g) Initial mask from color thresholding: $45 < R < 198$ and $G < 140$.

(h) Result of eroding the mask in (g), which breaks filamentous tape residue into smaller fragments.

(i) Regions in (h) after filtering by area, complexity, solidity, and extent — leaving three connected flake candidates.

(j) Bounding boxes from (i) overlaid on (f). Two flakes with appropriate contrast are selected for subflake segmentation. A region consisting of tape residue is also cropped but will be discarded later in subflake segmentation. A thick graphite flake (>20 nm), less useful for device fabrication, is excluded due to color filtering.

### Subflake Segmentation by Thickness Contrast

To identify internal thickness inhomogeneities, we perform subflake segmentation within each detected flake as the second step of our hierarchical pipeline. This is particularly important for applications requiring atomically uniform regions, such as high-mobility transport devices, where internal step edges act as scattering centers that degrade device performance. [7]

We apply k-Means clustering to the RGB values of each flake in the raw (non-devignetted) image after Gaussian blurring. On a 285 nm $SiO_2$/Si substrate, thin-film interference causes flake color to correlate strongly with thickness. The clustering process groups pixels of similar color into contiguous regions, which we interpret as subflakes—domains of approximately uniform thickness. This unsupervised approach is simple, interpretable, and robust across flakes with varying optical contrast and structural complexity. [13–15]

The number of color clusters ($k$) determines the granularity of the segmentation. For hBN flakes, which are typically thicker (up to tens of nanometers), the broader optical contrast range warrants a higher $k$ (typically $k$ = 21) to capture finer gradations. In contrast, graphene flakes usually span only a few atomic layers, with more subtle color variation, so a smaller $k$ (typically $k$ = 11) is sufficient to resolve meaningful internal structure without over-segmentation. While these defaults work well in most cases (Supplementary Information S4), certain flakes benefit from a different $k$, depending on their internal step complexity: too small a $k$ may obscure weak subdomains, whereas too large a $k$ can fragment thin flakes into noisy, incoherent clusters. (Supplementary Information S5) Future implementations may incorporate a data-driven model to adaptively choose flake-specific $k$ values.

Importantly, we adopt a **flake-first segmentation strategy**, performing k-Means clustering only within each isolated flake, rather than across the full image. This design offers critical advantages: the number of internal domains within a flake is small and predictable, enabling compact, physically meaningful $k$ values. In contrast, direct clustering on the full microscope image would require unpredictable and often excessive $k$, due to variable flake counts and background content, degrading both spatial coherence and interpretability. By constraining segmentation to individual flakes, we achieve robust and consistent subflake delineation, tailored to each flake's optical profile.

To avoid segmentation of regions without thickness variation, we apply a **post-processing merging step** based on RGB color distance. Specifically, adjacent clusters whose mean RGB values lie within a fixed Euclidean distance threshold are recursively merged. This merger is **transitive**: if cluster A is close to B, and B is close to C, then all three are merged.

For **hBN**, where thickness contrast is more gradual and subtle, we use a conservative threshold of **2** in RGB space (defined as the Euclidean norm):

$$\sqrt{(R_1 - R_2)^2 + (G_1 - G_2)^2 + (B_1 - B_2)^2} \tag{1}$$

For **graphene**, the typical RGB values on a 285 nm $SiO_2$ substrate are more widely separated adjacent thicknesses (bilayer vs monolayer, etc) can exceed **15**, so we adopt a more generous merging threshold of **4** to ensure that meaningful distinctions between

mono-, bi-, and few-layer graphene are preserved while still avoiding noise-driven over-segmentation. Thus, for most flakes, the segmentation result is robust with respect to the choice of *k* (Supplementary Info S4.)

Not all clusters produced by k-Means correspond to subflake regions. Some may contain background pixels, tape residue, or dirt particles. To filter out these non-flake regions, we process each cluster by examining its individual connected components. Only components that can bound a 5um x 5um square are retained. For each retained region, we compute the average RGB value from the de-vignetted image, and compare it against a pre-established RGB–thickness calibration curve (obtained via AFM). A component is accepted as a subflake only if its average color matches a known flake thickness within a specified RGB deviation threshold. In this process, most tape residues and dirt particles are effectively excluded. Their optical colors do not match any physically meaningful thickness values in our AFM-calibrated RGB–thickness mapping, and their connected components are often too small to pass the area threshold. As a result, these non-flake features are filtered out without requiring explicit classification, ensuring that only meaningful subflake regions are retained for downstream analysis.

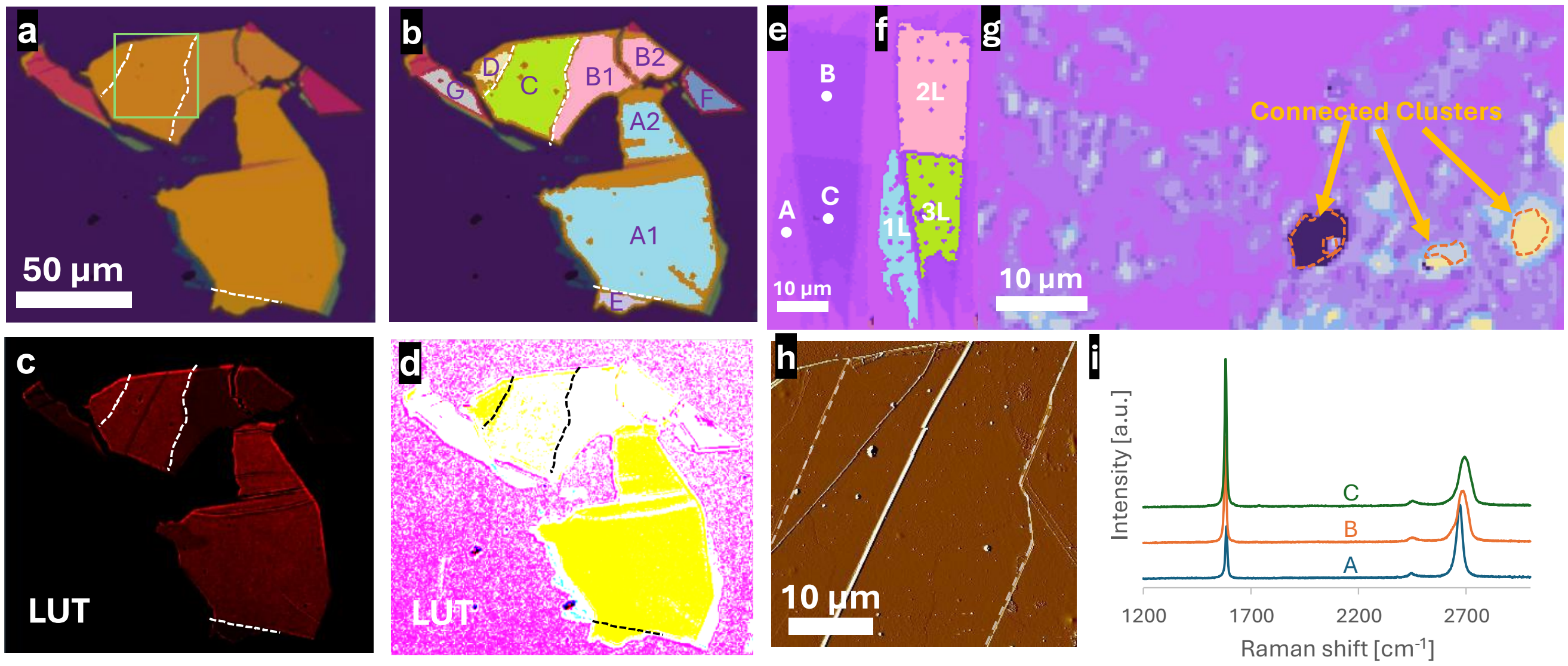

**Figure 3.** ***k-Means–based subflake segmentation of 2D material flakes.***

(a–b) Subflake segmentation results on the hBN flake from Figure 2a–e, based on subtle color variation. Dashed line A1–E highlights a visually apparent step edge separating two subflakes. Dashed lines B1–C and C–D indicate color boundaries that are not visible to the naked eye but are confirmed via (Look-Up Table) LUT-enhanced contrast (c-d, also see Supplementary Information S3) and AFM topography.

(e–f) Segmentation of the large thin graphene flake from Figure 2j. The segmented subregions correspond to monolayer, bilayer, and trilayer domains, independently verified by Raman spectroscopy (i) on points A–C.

(g) k-Means result on a tape residue region from Figure 2j. The fragmented clusters lack large connected regions with graphene-like colors and are correctly rejected by the post-segmentation filter.

(h) Non-contact mode (NCM) AFM amplitude (forward scan) image of the hBN flake, corresponding to the region highlighted by the green box in (a). Step edges indicated by white dashed lines in (a) are also marked here. The two additional features between these step edges correspond to wrinkles in the flake, rather than true thickness variations.

(i) Raman spectroscopy of the graphene sample shown in (e) at points A, B, and C, respectively. The evolution of the 2D peak shape and width confirms that the regions correspond to monolayer (A), bilayer (B), and trilayer (C) graphene.For more examples of k-Means based subflake segmentation, see Figure S5

### A structured flake database

To facilitate wafer-scale device planning, each detected flake and its corresponding subflakes are recorded in a structured database. Each **subflake** is represented by a row in a .csv file and uniquely indexed by [Image_x, Image_y, Flake ID, Subflake ID]. Ssubflakes sharing the same [Image_x, Image_y, Flake ID] belong to the same parent flake. (Fig. 4)

For each subflake, the system stores:

- its **average RGB value in the de-vignetted image** (used for thickness estimation via a pre-calibrated color-thickness mapping),
- its **area** in $\mu m^2$,
- the coordinates of the **largest rectangle fully contained** within the subflake (Inner Dimensions), and
- the coordinates of the **smallest rectangle that bounds the entire flake** (Outer Dimensions), shared by all subflakes that compose the flake.

This structured representation enables efficient filtering, sorting, and selection of flakes and subflakes based on thickness, area, aspect ratio, or position—streamlining downstream tasks such as stack design and layer matching.

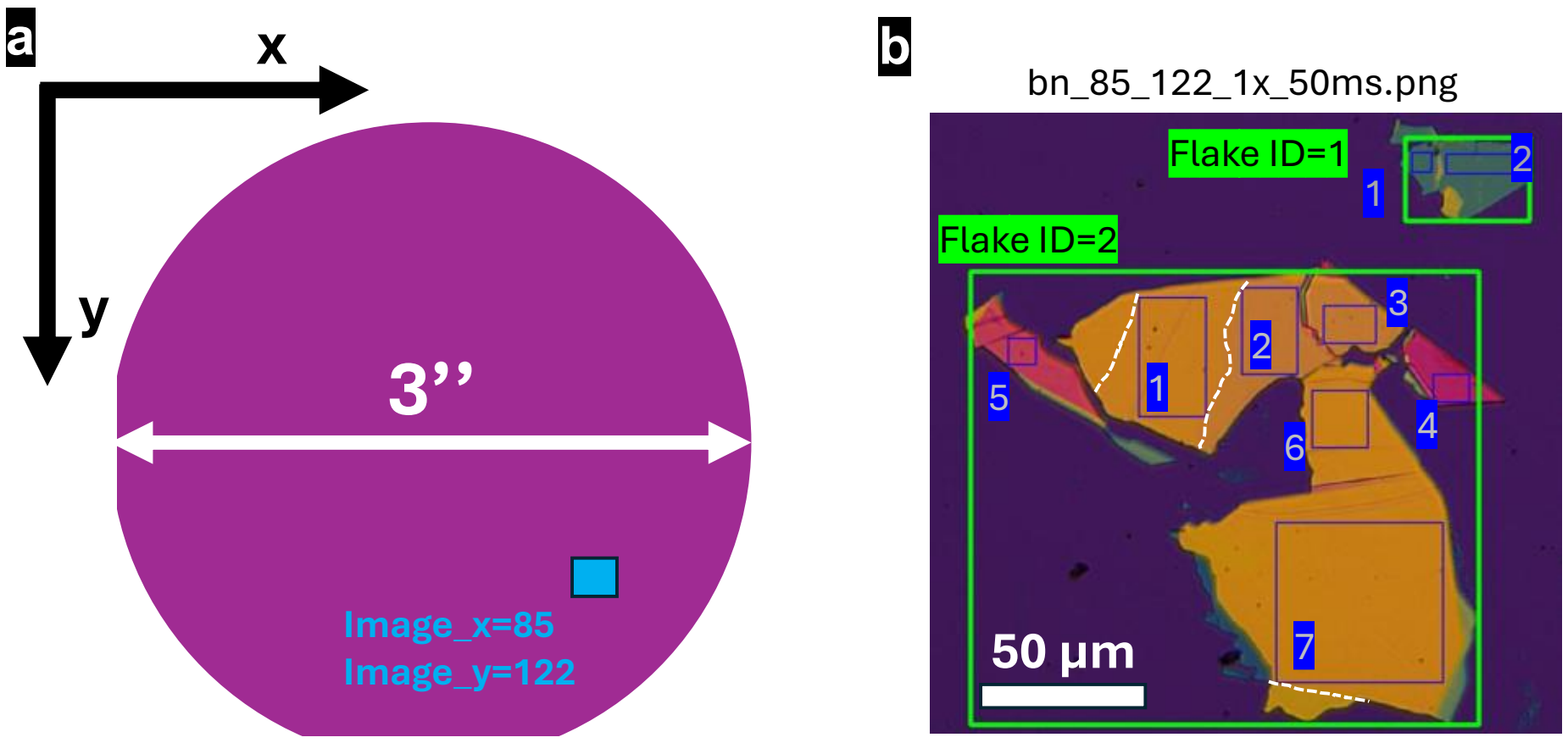

c

| Image_x | Image_y | Flake ID | Subflake ID | Area (μm$^2$) | R | G | B | Thickness (nm) | Inner_x1 (px) | Inner_x2 (px) | Inner_y1 (px) | Inner_y2 (px) | Outer_x1 (px) | Outer_x2 (px) | Outer_y1 (px) | Outer_y2 (px) |
|---|---|---|---|---|---|---|---|---|---|---|---|---|---|---|---|---|
| 85 | 122 | 1 | 1 | 66 | 61 | 94 | 86 | 21 | 685 | 697 | 530 | 542 | 681 | 757 | 521 | 572 |
| 85 | 122 | 1 | 2 | 347 | 60 | 94 | 86 | 21 | 706 | 750 | 531 | 543 | 681 | 757 | 521 | 572 |
| 85 | 122 | 2 | 1 | 1434 | 196 | 119 | 27 | 67 | 518 | 559 | 619 | 692 | 438 | 726 | 603 | 881 |
| 85 | 122 | 2 | 2 | 931 | 187 | 113 | 45 | 74 | 581 | 615 | 613 | 666 | 438 | 726 | 603 | 881 |
| 85 | 122 | 2 | 3 | 404 | 188 | 114 | 45 | 74 | 631 | 663 | 624 | 647 | 438 | 726 | 603 | 881 |
| 85 | 122 | 2 | 4 | 273 | 169 | 30 | 93 | 99 | 698 | 720 | 666 | 683 | 438 | 726 | 603 | 881 |
| 85 | 122 | 2 | 5 | 332 | 187 | 55 | 79 | 93 | 438 | 455 | 644 | 660 | 438 | 726 | 603 | 881 |
| 85 | 122 | 2 | 6 | 859 | 198 | 123 | 19 | 68 | 624 | 658 | 676 | 711 | 438 | 726 | 603 | 881 |
| 85 | 122 | 2 | 7 | 3652 | 198 | 124 | 18 | 68 | 602 | 704 | 757 | 855 | 438 | 726 | 603 | 881 |

**Figure 4. *Flake database.***

(a) Schematics of the wafer coordinate system as defined by the crystallographic axes of Si. [Image_x, Image_y] locates a specific microscope image scanned on the wafer.

(b) The result of flake and subflake segmentations on the hBN flake shown in Fig. 3a. Outer Dimensions of the flakes are shown in green boxes, while the Inner Dimensions of each subflake are shown in blue boxes. White dashed line shows internal step edges that are hard to identify.

(c) Entries in the flake database that are associated with the 2 flakes, 9 subflakes shown in (b). Each subflake corresponds to one entry. Notice that subflakes that belong to the same flake share the same Outer Dimensions.

We applied this system to a full 3-inch $SiO_2$/Si wafer, mechanically exfoliated with graphene on one half and hBN on the other. The entire pipeline—from wafer alignment to image acquisition and flake analysis—completed in under 12 hours. In total, over **6,000 valid subflakes** were identified across the wafer. Over 99% of the segmented subflakes correspond to actual 2D material features, rather than spurious artifacts such as **tape residue, polymer contamination, or dust particles**.

For graphene flakes, the average RGB values of subflakes follow distinct trajectories in R–G color space and form discrete clusters corresponding to 1–8 atomic layers. This enables rapid and reliable identification of monolayer and few-layer regions (see Fig. 6), offering a practical basis for automated layer selection in downstream stacking. [16–18]

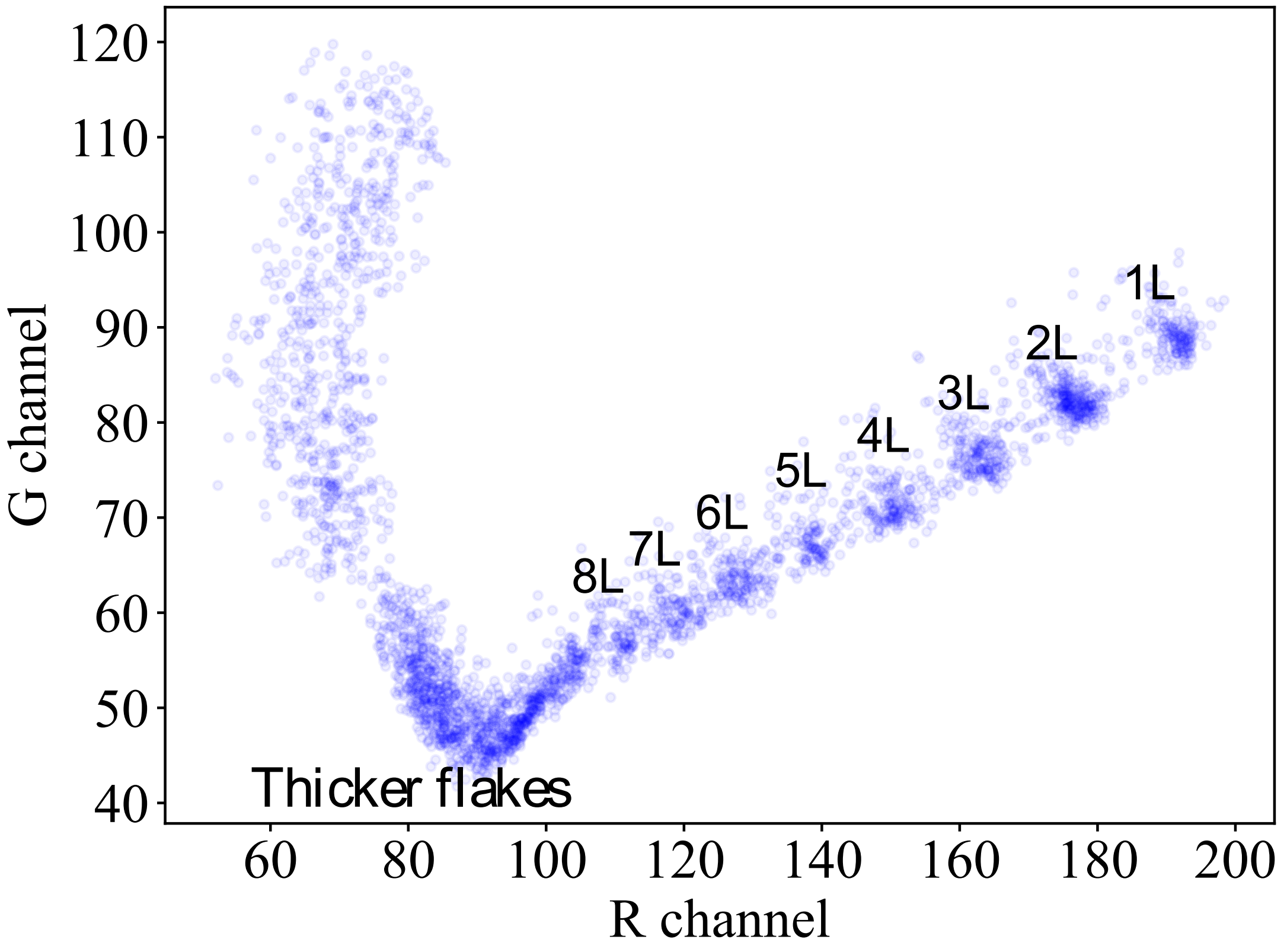

**Figure 5. RGB-based clustering of graphene flakes by thickness.**
Average R and G channel values for 3,112 graphene subflakes extracted from the wafer. Each point represents a subflake, and the labeled clusters correspond to regions identified as 1 to 8 graphene layers based on optical color. The clusters are disjoint and well-separated in R–G space, reflecting the discrete nature of optical contrast on 285 nm $SiO_2$/Si substrate. This enables rapid and reliable thickness assignment without nanoscale measurements such as AFM or Raman. Flakes thicker than 8 layers follow a different trajectory due to color reversal in thin-film interference. This separation validates the effectiveness of our segmentation and indexing pipeline for layer-selective device design.

To support downstream heterostructure assembly, we developed two companion tools, Stack Designer (Fig. 6) and Flake Gallery (Fig. S7), which provide a graphical interface to browse the flake library, select compatible subflakes based on thickness and shape, and plan multi-layer pickup sequences. Each selected stack blueprint can be exported with precise relative alignment parameters, forming a bridge from flake detection to automated flake stacking. [19–23]

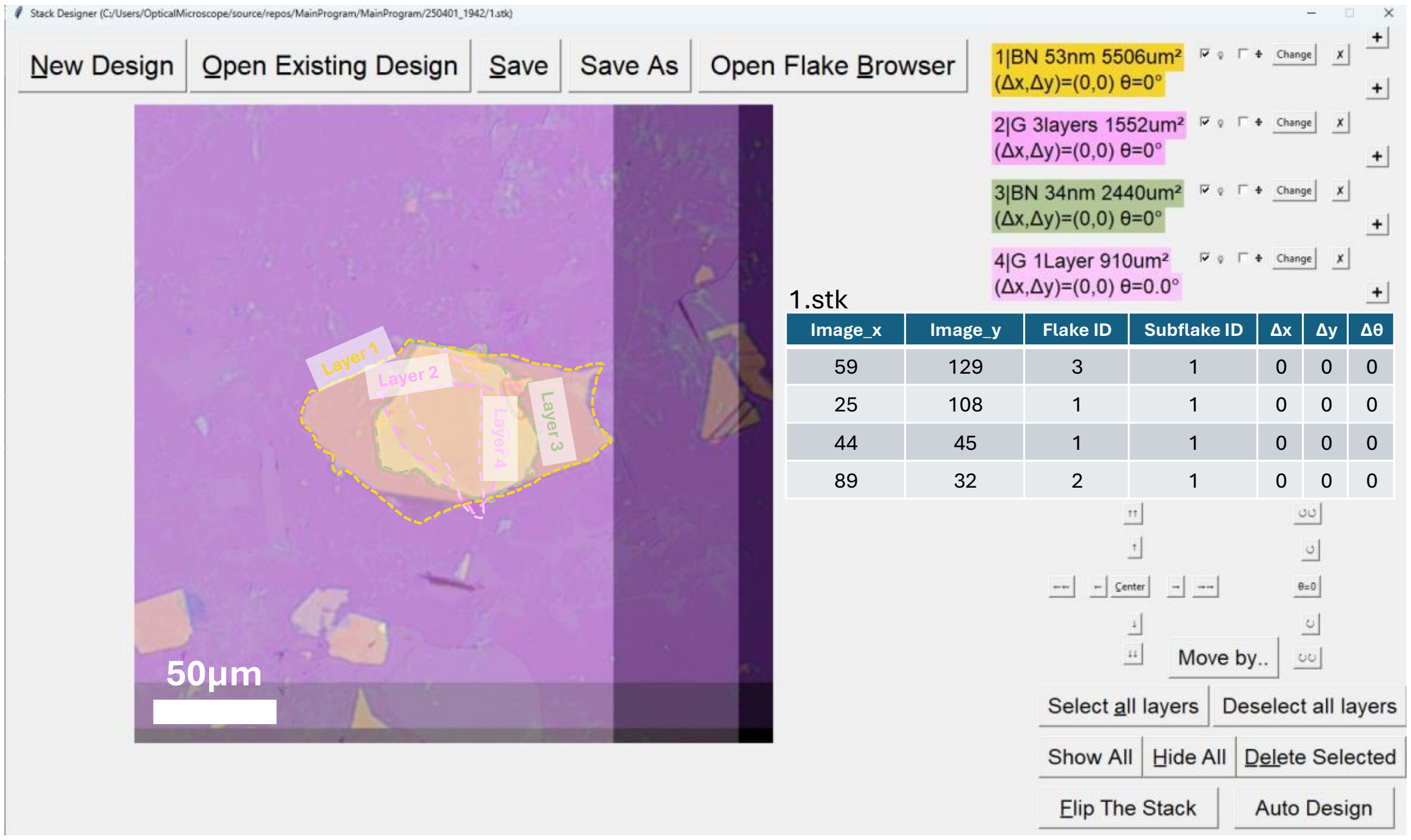

**Figure 6 | Stack Designer Graphical User Interface (GUI).**
The Stack Designer, used in conjunction with the Flake Gallery (Fig. S7), enables creation of .stk files that serve as blueprints for stacking. Each row specifies the subflake ID and its positional offsets (Δx, Δy, Δθ). When all offsets are set to zero, the subflake is positioned such that the center of its Inner Dimensions aligns with the microscope field center. In this default configuration, the core regions of all layers in the stack exhibit near-maximal overlap.

## Conclusion

We have developed a cost-effective (< $35 k hardware) and scalable image-processing pipeline for automated detection, characterization, and indexing of exfoliated 2D material flakes. By combining a CNN-based material classifier with a transparent, two-stage segmentation strategy, the system robustly processes BN and graphene flakes under optimized illumination conditions. Material categorization is performed via supervised learning, while flake and subflake segmentation rely on interpretable, human-like operations (edge- and color-based detection followed by K-means clustering). The

approach minimizes the need for extensive ground-truth datasets—requiring only limited calibration, such as RGB–thickness mapping—while maintaining physical interpretability throughout the pipeline.

Integrated with blueprint-based stacking tools, the workflow forms a direct bridge from wafer-scale image acquisition to deterministic 2D layer assembly. This approach lowers the barrier to automation in van der Waals heterostructure fabrication, enabling smaller laboratories to deploy high-throughput stacking capabilities previously limited to heavily customized, high-cost platforms. Looking forward, coupling such pipelines with automated pick-up hardware, adaptive alignment algorithms, and database-driven layer selection could enable fully autonomous heterostructure assembly—accelerating materials discovery, improving device reproducibility, and expanding access to 2D materials research worldwide. [3,19–21,23,24]

**References**

1. Novoselov, K. S., et al. Electric field effect in atomically thin carbon films. Science 306, 666–669 (2004). https://doi.org/10.1126/science.1102896

2. Splendiani, A., et al. Emerging photoluminescence in monolayer MoS2. Nano Letters 10, 1271–1275 (2010). https://doi.org/10.1021/nl903868w

3. Geim, A. K. & Grigorieva, I. V. Van der Waals heterostructures. Nature 499, 419–425 (2013). https://doi.org/10.1038/nature12385

4. Novoselov, K. S., et al. 2D materials and van der Waals heterostructures. Science 353, aac9439 (2016). https://doi.org/10.1126/science.aac9439

5. Manzeli, S., et al. 2D transition metal dichalcogenides. Nature Reviews Materials 2, 17033 (2017). https://doi.org/10.1038/natrevmats.2017.33

6. Liu, Y., et al. Van der Waals heterostructures and devices. Nature Reviews Materials 1, 16042 (2016). https://doi.org/10.1038/natrevmats.2016.42

7. Dean, C. R., et al. Boron nitride substrates for high-quality graphene electronics. Nature Nanotechnology 5, 722–726 (2010). https://doi.org/10.1038/nnano.2010.172

8. Butler, S. Z., et al. Progress, challenges, and opportunities in two-dimensional materials beyond graphene. ACS Nano 7, 2898–2926 (2013). https://doi.org/10.1021/nn400280c

9. Wang, Q. H., et al. Electronics and optoelectronics of two-dimensional transition metal dichalcogenides. Nature Nanotechnology 7, 699–712 (2012). https://doi.org/10.1038/nnano.2012.193

10. Lin, X., et al. Intelligent identification of two-dimensional nanostructures by machine-learning optical microscopy. Nano Research 11, 6311–6322 (2018). https://doi.org/10.1007/s12274-018-2155-0

11. Masubuchi, S., et al. Classifying optical microscope images of exfoliated graphene flakes by data-driven machine learning. npj 2D Materials and Applications 3, 4 (2019). https://doi.org/10.1038/s41699-018-0084-0

12. Sanchez-Juarez, J., et al. Automated system for the detection of 2D materials using digital image processing and deep learning. Optical Materials Express 12, 1214–1232 (2022). https://doi.org/10.1364/OME.448523

13. Sterbentz, R. M., et al. Universal image segmentation for optical identification of 2D materials. Scientific Reports 11, 5934 (2021). https://doi.org/10.1038/s41598-021-85159-9

14. Blake, P., et al. Making graphene visible. Applied Physics Letters 91, 063124 (2007). https://doi.org/10.1063/1.2768624

15. Ni, Z., et al. Graphene thickness determination using reflection and contrast spectroscopy. Nano Letters 7, 2758–2763 (2007). https://doi.org/10.1021/nl071254m

16. Jessen, B. S., et al. Quantitative optical mapping of two-dimensional materials. Scientific Reports 8, 6381 (2018). https://doi.org/10.1038/s41598-018-23922-1

17. Li, H., et al. Rapid and reliable thickness identification of two-dimensional nanosheets using optical microscopy. ACS Nano 7, 10344–10353 (2013). https://doi.org/10.1021/nn4047474

18. Mondal, M., et al. Optical microscope based universal parameter for identifying layer number in two-dimensional materials. ACS Nano 16, 12933–12941 (2022). https://doi.org/10.1021/acsnano.2c04833

19. Masubuchi, S., et al. Autonomous robotic searching and assembly of two-dimensional crystals to build van der Waals superlattices. Nature Communications 9, 1413 (2018). https://doi.org/10.1038/s41467-018-03723-w

20. Mannix, A. J., et al. Robotic four-dimensional pixel assembly of van der Waals solids. Nature Nanotechnology 17, 361–366 (2022). https://doi.org/10.1038/s41565-021-01061-5

21. Zomer, P. J., et al. Fast pick up technique for high-quality heterostructures of bilayer graphene and hexagonal boron nitride. Applied Physics Letters 105, 013101 (2014). https://doi.org/10.1063/1.4886096

22. Forsythe, C. M., et al. Band structure engineering of 2D materials using patterned dielectric superlattices. Nature Nanotechnology 13, 566–571 (2018). https://doi.org/10.1038/s41565-018-0138-7

23. McGilly, L. J., et al. Visualization of moiré superlattices. Nature Nanotechnology 15, 580–584 (2020). https://doi.org/10.1038/s41565-020-0708-3

24. Lei, Y., et al. Graphene and Beyond: Recent Advances in Two-Dimensional Materials Synthesis, Properties, and Devices. ACS Nanoscience Au 2, 112–142 (2022). https://doi.org/10.1021/acsnanoscienceau.2c00017

25. Wang, Q. H., et al. Electronics and optoelectronics of two-dimensional transition metal dichalcogenides. Nature Nanotechnology 7, 699–712 (2012). https://doi.org/10.1038/nnano.2012.193

**Table of Contents**

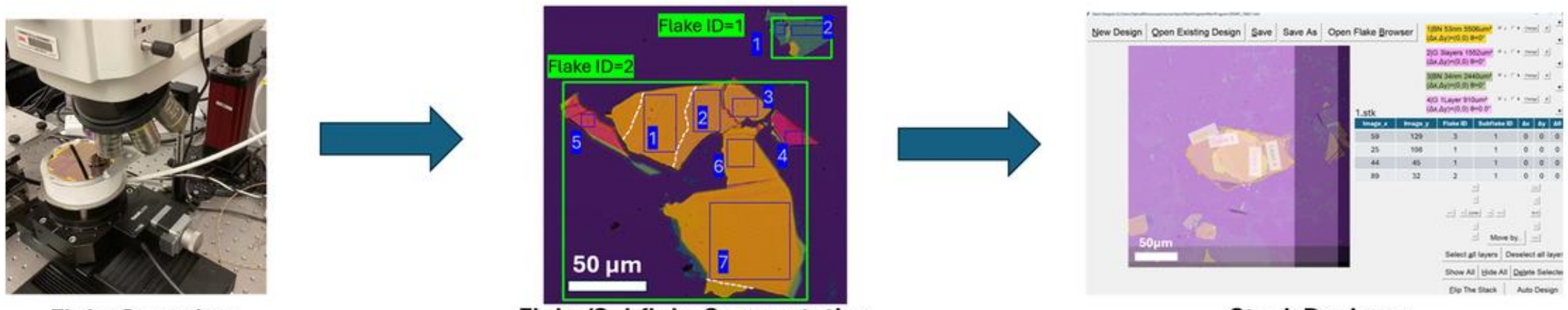

Flake Scanning    Flake/Subflake Segmentation    Stack Designer

This article presents a robust, cost-effective image-processing pipeline for two-dimensional materials. Beyond flake segmentation, the method resolves optical step edges when contrast permits, enabling thickness-aware indexing. By exporting structured flake information directly usable in robotic workflows, the system integrates seamlessly into automated stacking platforms, lowering the barrier for scalable van der Waals heterostructure assembly.

**Acknowledgements**

This study is supported by the Office of Basic Energy Sciences, Materials Science and Engineering Division, US Department of Energy under contract no. DE-SC0012704 and the Department of Energy (DOE) ’s Energy I-Corps Program Cohort 20 by DOE Office of Science, and Advanced Materials and Manufacturing Technologies Office (AMMTO).

Raman spectroscopy measurements are performed at Columbia Nano Initiative (CNI) of Columbia University.

We acknowledge fruitful discussions with Dr. Raymond Blackwell and Mike Clarkin, and their participation in the Energy I-Corps Program.

**Conflict of Interest Disclosure**

Y.L. discloses a preliminary patent application related to the AutoLab system, submitted through Brookhaven National Laboratory (BNL). The other authors declare no competing interests.